# An Extensive Study of Residential Proxies in China

Mingshuo Yang[*]
Shandong University
China
yangmingshuo@mail.sdu.edu.cn

Yunnan Yu[*]
University at Buffalo
United States
yunnanyu@buffalo.edu

Xianghang Mi[†]
University of Science and
Technology of China
China
xmi@ustc.edu.cn

Shujun Tang
QI-ANXIN Technology Research
Institute
China
tangshujun@qianxin.com

Shanqing Guo[†‡]
Shandong University
China
guoshanqing@sdu.edu.cn

Yilin Li
Shandong University
China
liyilin22@mail.sdu.edu.cn

Xiaofeng Zheng[§]
Tsinghua University
China
zxf19@mails.tsinghua.edu.cn

Haixin Duan[§]
Tsinghua University
China
duanhx@tsinghua.edu.cn

## Abstract

We carry out the first in-depth characterization of residential proxies (RESIPs) in China, for which little is studied in previous works. Our study is made possible through a semantic-based classifier to automatically capture RESIP services. In addition to the classifier, new techniques have also been identified to capture RESIPs without interacting with and relaying traffic through RESIP services, which can significantly lower the cost and thus allow a continuous monitoring of RESIPs. Our RESIP service classifier has achieved a good performance with a recall of 99.7% and a precision of 97.6% in 10-fold cross validation. Applying the classifier has identified 399 RESIP services, a much larger set compared to 38 RESIP services collected in all previous works. Our effort of RESIP capturing lead to a collection of 9,077,278 RESIP IPs (51.36% are located in China), 96.70% of which are not covered in publicly available RESIP datasets. An extensive measurement on RESIPs and their services has uncovered a set of interesting findings as well as several security implications. Especially, 80.05% RESIP IPs located in China have sourced at least one malicious traffic flows during 2021, resulting in 52-million malicious traffic flows in total. And RESIPs have also been observed in corporation networks of 559 sensitive organizations including government agencies, education institutions and enterprises. Also, 3,232,698 China RESIP IPs have opened at least one TCP/UDP ports for accepting relaying requests, which incurs non-negligible security risks to the local network of RESIPs. Besides, 91% China RESIP IPs are of a lifetime less than 10 days while most China RESIP services show up a crest-trough pattern in terms of the daily active RESIPs across time.

---
[*]Both authors contributed equally to this research.
[†]Corresponding authors.
[‡]Also with Quancheng Laboratory.
[§]Also with QI-ANXIN Technology Research Institute.

## 1 Introduction

In a recent cyberattack campaign attempting to break into hundreds of intelligence targets in US during 2021, residential proxies (RESIPs) were used by attackers to masquerade themselves as benign Americans [43]. These attacks are believed to be ongoing with intended victims including US government agencies, non-government organizations, as well as critical IT firms. Since residential proxies are widely located in home networks or cellular networks, traffic relayed through residential proxies will look much less suspicious compared to traditional proxies such as VPN and Tor. This characteristic is being utilized by attackers to masquerade their attack traffic as benign visits from employees and thus evade detection or blocking deployed by organizations. Also, residential proxy services are public available on the Internet and can be easily purchased with few or no identity checking [25].

As RESIP services are becoming increasingly popular since emerged in 2016, it has attracted not only abuse from attackers, but also interests from the research community. [25] conducted a large-scale infiltration on 5 representative RESIP services with more than 6-million RESIP IPs identified. Then [44] steps forward to reveal that RESIP proxies may be harvested through unauthorized NAT entry injection by exploiting a UPNP vulnerability. Besides, [26] moves the spotlight closer to mobile devices and explores how mobile devices have been recruited to serve as RESIPs along with a set of mobile proxy SDKs identified and profiled.

However, the RESIP ecosystem is quickly evolving, e.g., by December 2021, among 38 RESIP services studied in these previous works, 7 have disappeared and one (Luminati) has rebranded itself as BrightData [23]. Also, RESIP services in previous works were identified either manually or semi-automatically, and it turns out that their coverage of RESIP services is low, as demonstrated in our study (§3.1). Therefore, it is unclear whether the understandings generated from previous works can still be applicable to the up-to-date and ever-evolving RESIP ecosystem. Besides, each of previous works profiled RESIP services for only several months, due to the costly infiltration framework which requires service purchase and traffic relaying. For such a ever-growing ecosystem, an low-cost methodology is necessary to continuously capture RESIPs and profile RESIP services. Further, previous works are dedicated to RESIP services available on the English Internet, and have missed region-dedicated RESIP services. Especially, our study has identified 64 previously unknown RESIP services that are operated in China and dedicated to the China market. Many of these Chinese RESIP services have been there for years and almost claim to have millions of RESIPs located in China. Although previous works have captured globally distributed RESIPs at the scale of millions, very few of them are located in China. In summary, little is known about this regional RESIP ecosystem as well as its security implications. Lastly, although [26] has revealed the recruitment of mobile devices by several RESIP services, the supply chain of RESIPs has yet to be further uncovered, which is crucial for understanding and addressing the security risks of RESIPs.

Both the ever-evolving RESIP ecosystem and the aforementioned limitations in previous works motivate us to carry out this study. Standing on the shoulders of previous works, our study aims to move forward in the following directions. First of all, we want to achieve a in-depth understanding of the RESIP ecosystem in China with a focus on its landscape, evolution, and security implications. Besides, we want to build up a detector that can automatically identify all public available RESIP services regardless of regions or languages. Furthermore, we want to identify low-cost techniques for capturing RESIPs and build up a radar to continuously identify and profile newly emerged RESIPs. Lastly, we want to dig deeper in the supply chain of RESIPs and uncover any previous unknown participants in this ecosystem.

To achieve these research goals, we first explore how to automatically identify public available RESIP services. We observe that a RESIP service (RPS) usually operates an independent website for service promotion and customer management. Also, a RPS website has unique semantic features especially on the landing page. Particularly, a typical RPS website tend to display RPS-relevant keywords of various categories in the text elements including the title, the description, among others. Typical keywords relevant to RPS can be proxy IP types (e.g., *residential IP* and *mobile proxy*), proxy protocols (e.g., *SOCKS*, *HTTPS*), and proxy features (e.g., *unblockable* and *undetectable*). Leveraging these observations, we engineer a set of robust features and build up an effective classifier to automatically decide whether a given website is RPS or not. However, One obstacle we encountered is the unavailability of enough RPS websites for training and evaluation, since previous works identified only 38 RPS websites in total and 7 of them have disappeared. We addressed this by means of a strategy involving multiple iterations of training weak classifiers, predicting, and manually verifying sampled positive predictions to extend the ground truth. Our ultimate RPS classifier has achieved a recall of 99.53% and a precision of 97.62% under 10-fold cross validation. We then applied this classifier to identifying Chinese and English RPS websites. In total, 399 RPS websites have been captured, among which, 368 are unknown before. These newly identified RPS websites include 64 Chinese RESIP services and 304 English RESIP services.

Given the 64 Chinese RESIP services, we move to carry out the first in-depth characterization of this ecosystem, starting from capturing RESIPs. We first selected out 5 Chinese RESIP services and carried out a 6-month infiltration by adopting the framework proposed in [25]. The idea behind this framework is to relay well-crafted HTTP/HTTPs requests through the RESIP services, to web servers under our control, which will allow the web servers to observe IP addresses of the exit nodes, namely, the RESIPs. However, as discussed above, such a infiltration is costly and involves much manual effort for every RESIP service. This costly infiltration framework is necessary for traditional RESIPs which work in a backconnect mode and RESIPS are hidden behind the gateway servers controlled by the RESIP service. However, we observe that many RESIP services as identified by our classifier offer RESIPs without gateway servers as the intermediate. Specifically, RESIPs in this new category will listen to specific TCP/UDP ports and accept relaying requests that are directly originated from the proxy customers. To distinguish from traditional backconnect RESIPs (BC-RESIPs), we call such RESIPS as direct RESIPs. Also observed is that some proxy services will bind active direct RESIPS to their subdomains through DNS. Among all Chinese RESIP services, 42 have been confirmed to offer RESIPs through such a channel. Leveraging Passive DNS, we are allowed to extract all the historical RESIPs along with their lifetime. We call these direct RESIPs as DP-RESIPs since they are captured through passive DNS. Another channel we utilized to capture RESIP is RESTful APIs exposed by some RESIP services. These RESTful APIs are provided to proxy customers to fetch up-to-date direct RESIPs. To distinguish from BC-RESIPs and DP-RESIPs, we call such RESIPs as DA-RESIPs since they are collected through APIs. In total, we have captured 9,077,278 RESIP IPs among which, 8,176,522 are BC-RESIP IPs, 2,283,665 are DA-RESIP IPs, and 1,471,361 are DP-RESIP IPs.

We then carried out an extensive measurement on the captured RESIP services and RESIP IPs, which is facilitated by two threat intelligence platforms along with several other complimentary datasets. The key findings of our study are listed below.

● We have captured the largest RESIP dataset with 9,077,278 RESIP IPs and 399 RESIP services. More importantly, 64 RESIP services are dedicated to China while 4,661,934 RESIP IPs are located in China, which is much larger than previous datasets. In addition to a large volume of RESIP IPs in China, the left RESIP IPs are still widely distributed in 226 countries and regions, 49,461 ISPs, 207 /8 IPv4 CIDRs, and 24,691 autonomous systems, which is comparable to previous RESIP datasets. Also, compared with RESIP datasets revealed in previous works, 8,777,808 RESIP IPs and 368 RESIP websites are newly captured.

● Leveraging passive DNS, we are allowed to profile for the first time the evolution and lifetime of RESIPs and RESIP services. Among the 42 RESIP services offering DP-RESIPs, many show up a crest-trough pattern for the number of daily active RESIPs across time. And it usually takes 3 years on average for a RESIP service to arrive at the crest, in another word, gain the maximum number of RESIPs. On the other hand, a RESIP service can collapse quickly in less than 129 days from the crest to the trough. Regarding RESIPs,

RESIPs are observed to have a short lifetime, regardless of RESIP services, e.g., 91% DP-RESIPs have a lifetime of less than 10 days.

● Leveraging multiple threat intelligence platforms, we have identified various malicious activities colocated with RESIPs during the same period. Especially, 62.61% China RESIP IPs have been detected to conduct cryptojacking activities with each associated with an average number of 36 cryptojacking traffic flows during 2021. Also, 3129 China RESIP IPs are found to have ever distributed payloads of a large-scale botnet named Mozi, while 21.00% China RESIP IPs were detected as bots by a proprietary threat intelligence platform.

● China RESIPs have been observed in corporation networks of 559 sensitive organizations including government agencies, education institutions and enterprises. One explanation is that some devices in the local networks of these organizations serve as RESIPs, which raises concerns regarding the security of local network of these organizations. Also, each of the aforementioned 3,232,698 direct China RESIPs have opened at least one TCP/UDP ports as instructed by the respective RESIP service, which also raises non-negligible potential risks to the local networks and the local devices.

In summary, we have carried out an in-depth characterization of the RESIP ecosystem in China, with a set of insightful findings and observations. Also, our study is made possible through a novel RPS website classifier as well as multiple new channels to collect RESIP IPs in an continuous and low-cost manner, which, along with the resulting RESIP datasets, can benefit future research in this area. The resulting datasets are made publicly available on https://rpaas.site.

## 2 Background

**Residential proxies**. In recent years, residential proxies (RESIPs) keep evolving with more service providers emerging and ever-growing proxy scale. A typical RESIP works in a backconnect mode wherein the RESIP node is hidden behind the gateway servers operated by the respective service. In the backconnect mode, proxy traffic sourced from a proxy customer will be first routed to the gateway server, which in turn forwards it to the specific RESIP node, then to the traffic destinations. What's more, most services will allow proxy customers to specify where they want to exit the relayed traffic, in terms of countries and cities. Also, proxy customers are able to stick their traffic to the same exit node, either through passing the same session id, or connect to some sticky proxy gateways which usually bind to a exit node for every 5 or 10 minutes [35]. Varied by RESIP services, multiple proxy protocols are supported, especially HTTP/HTTPS and SOCKS. A new trend we have observed is the support of more proxy protocols by RESIP services, compared to previous works. Specifically, as revealed in previous works, RESIP services usually support only HTTP/HTTPS and SOCKS protocols. However, by looking up documents offered by the 47 China RESIP services under our study, we find out that 34 of them have also supported at least one VPN protocols including PPTP, L2TP , SSTP, OpenVPN, and IPsec. Regarding pricing, a RESIP service usually offers several monthly subscription options which differ in the price and resource constraints (e.g., traffic volume). Taking ProxyRack, one of the long-existing and popular RESIP service, as an instance, a monthly subscription with 10GB data is offered at a price of 49.95\$[1], through which, proxy customers can access more than 5-million RESIP IPs. In the meantime, a monthly plan at 65.95\$ is also provided and it allows unlimited data access but is constrained with 5 concurrent connections[2].

In our study, the terms of *RESIP* and *RESIP IP* will be exchangeable, since the IP address is the only public available information we can have to refer to a RESIP. Although different RESIP services can be operated by the same underlying operator, only a few cases have been observed and it is challenging to accurately cluster all services into their respective operators. Therefore, we consider each RPS website as a RPS service that is independent from others unless deterministic connections have been observed.

**Passive DNS**. Passive DNS refers to datasets of historical DNS records. And each passive DNS record may also contain the cumulative query volume and the time frame from when the DNS record is first observed to when it expires. Passive DNS is usually collected from widely distributed DNS resolvers across a long period. Passive DNS serves an important role in security investigations such as detecting and profiling malicious domains [5, 6]. In our study, we apply passive DNS to identifying RESIPs as well as profiling evolution of RESIP services.

## 3 Methodology and Datasets

To profile the RESIP ecosystem in China, we need to first extensively identify RESIP services as well as selecting out representative ones. Once RESIP services identified and selected, their residential proxies (RESIPs) should be captured at a large scale and across a period long enough. In addition, relevant datasets (e.g., passive DNS) should also be collected to facilitate an in-depth characterization of RESIP services, RESIP devices, and RESIP IPs. This section serves to detail the respective methodologies as well as summarizing the resulting datasets.

### 3.1 Identifying RESIP Services

In previous works, RESIP services were collected either manually or semi-automatically, which is not suitable to continuously capture emerging RESIP services. Also, as demonstrated by our detection, previous works have unfortunately missed many RESIP services, regardless of the regions. In this study, we explore for the first time an intelligent pipeline to automatically identify RESIP services.

We observe that a RESIP service (RPS) usually operates an independent website for service promotion and customer management. Therefore, you can decide whether a website is a RPS by looking into the text elements of its website especially the website landing page (homepage). To automatically identify RESIP services that are public available on the web, we design and implement an NLP-driven website classifier to reliably decide whether a give website is a RPS, and we call this classifier as the residential proxy service classifier (RPSC).

---

[1] https://www.proxyrack.com/premium-geo-residential/#pricing
[2] https://www.proxyrack.com/private-residential-proxies/#pricing

**RPS candidates**. As the first step of our pipeline, we queried major search engines with RPS-relevant keywords. Among these well-crafted keywords, some are adopted from previous works [25, 26, 44] while others are manually extracted through visiting RPS websites. Table 1 lists them along with their translations in Chinese. These involved search engines include Google Search, Bing Search and Baidu Search. For each search keyword and its language variants, we query all search engines and retrieve up to 1K resulting entries for each query. In total, 12,591 distinct search result entries are collected with each uniquely identified by an URL. These 12,591 URLs belong to 4,675 distinct apex domains, each of which is considered as a RPS candidate. To automatically classify if an apex domain is an RPS website, we move to build up an effective classifier leveraging semantic features extracted from the webpages hosted under each apex domain.

**Groundtruth collection**. We then collected a groundtruth dataset consisting of both RPS websites and non-RPS ones. Specifically, the non-RPS websites were collected by selecting the top 1K out of the Tranco top website list [34]. However, it turns out to be challenging to collect a comparable amount of RPS cases since previous works [25, 26] revealed only 38 RPS websites in total, among which, 7 were found to be out of service by Nov 2021. To further increase RPS cases in our groundtruth, we carried out a collection process involving a cycle of training a weak classifier on the preliminary groundtruth dataset , applying the weak classifier to RPS candidates, sampling and verifying positive predictions, and extending the groundtruth with confirmed RPS predictions. And this cycle continues until a reasonable number of RPS cases are captured. In the training step, a preliminary classifier is trained upon the current groundtruth dataset. The resulting classifier, despite weak, can still help exclude non-RPS cases from the RPS candidates and thus narrow the search space for our manual confirmation. As the step of manual confirmation, we sample 50 positive predictions and manually decide whether there is true positive. A website is considered as a true positive (an authentic RESIP website) only when it has an independent brand name, and is offering proxy services of which the underlying proxies should be well claimed as residential ones. The confirmed true RPS cases are added to the groundtruth before entering the next iteration. Through 3 iterations, we have successfully collected 79 extra RPS websites. Finally, we have composed a groundtruth dataset consisting of 110 RPS websites and 1,073 non-RPS websites. This dataset, albeit not well balanced, is found to be good enough to train an effective RPS classifier, as evaluated below.

**Data collection and preprocessing.** Regardless of training the preliminary or the final classifiers, a prior step is to crawl and preprocess webpages for websites in the groundtruth dataset as well as all the RPS candidates. For each website, starting from the homepage. our crawler recursively visits any new in-site webpages as referred by the visited ones until the predefined threshold of 100 is reached or all the in-site webpages have been crawled. Also, considering web elements can be asynchronously loaded, instead of static crawling, our crawler dynamically renders each webpage by instructing a headless browser through the Selenium framework [37]. As a result, a webpage's HTML file, images, and JavaScript files will be loaded and saved as $data_{webpage}$.

**Table 1: RPS-relevant keywords for querying search engines.**

| No | English | Chinese |
| --- | --- | --- |
| 1 | residential ip provider | 住宅 IP 供应商 |
| 2 | residential proxy provider | 住宅代理供应商 |
| 3 | residential proxy service | 住宅代理服务 |
| 4 | residential proxies pricing | 住宅代理定价 |
| 5 | static residential ip | 静态住宅 IP |
| 6 | static residential proxy | 静态住宅代理 |
| 7 | dynamic ip proxy | 动态 IP 代理 |
| 8 | isp proxy | ISP 代理 |
| 9 | resi proxy | RESI 代理 |
| 10 | rotating proxy | 旋转代理 |
| 11 | proxy pool buy | 代理池购买 |
| 12 | buy proxy servers | 购买代理服务 |
| 13 | buy unlimited proxies | 购买无限代理 |

Once crawled, all the texts residing in the HTML DOM tree will be further extracted and grouped into 5 categories: the title, the description meta, the keywords meta, the tags meta, as well as the body text (text paragraphs in the body tag). Since a captured webpage can in various natural languages not limited to Chinese and English, Google Cloud Translation API is utilized to decide the corresponding natural language and translate the text paragraphs into English if necessary. Then, all these text paragraphs will further go through a set of well-adopted NLP preprocessing steps including tokenization, stop word removing, and lemmatization, before serving as the source to extract classification features.

**Feature engineering** Given the groundtruth dataset, we move forward to craft effective features from processed text paragraphs for each website. The features are designed to well distinguish RPS and non-RPS websites. We observe that a RPS website tend to present, especially on the homepage, RPS-specific keywords of various categories, such as proxy IP types (*residential IP* and *mobile proxy*), proxy protocol (*SOCKS*, *HTTPS*, and *UDP*), proxy features (*unblockable* and *undetectable*), purchase/subscription (*buy proxies* and *proxy access*). Therefore, we adopt a keyword-based feature engineering strategy and limit our feature engineering scope to the homepage of each website. In short, we have identified a set of 12 keywords (Table 8 in Appendix A), which are found to be effective in terms of distinguishing RPS and non-RPS websites. To identify these keywords, text paragraphs of each homepage is considered as a document. And tf-idf [36] is utilized to select out words that are important for RPS documents but not that important for non-RPS documents, more details can be found in Appendix A.

**Training and evaluation.** Upon groundtruth collected and features defined, we move to train and evaluate our classifiers. When training the preliminary classifiers, we hardcode the classification algorithm as Random Forest with 25 decision trees, in the hope of speeding up the extending of groundtruth. However, when training the final classifier, we experimented in details various classification algorithms along with their hyper-parameter combinations. Evaluation reveals that the best performance is achieved by Random Forest with 200 decision trees. And it achieves a recall of 99.72%, precision of 97.63%, and F-1 score of 98.66% in 10-fold cross validation. This model is thus considered as the final model to do predictions on the RPS candidates.

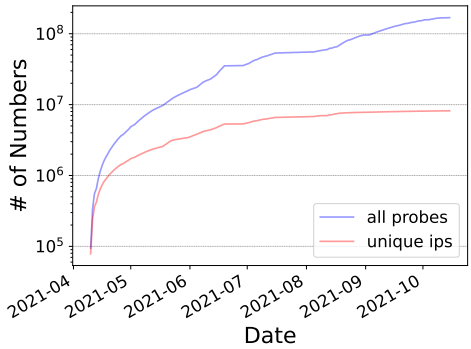

**Figure 1: Stats of our infiltration of China RESIP services in terms of cumulative probes and unique IPs.**

**Prediction and validation.** The final classification model has predicted 338 RPS candidates as positive (RPS) and the left as negative (non-RPS). To further validate the performance of this classifier, we manually visited each of these positive predictions to verify if it is a RPS website. It turns out 289 of them (85.50%) were found to be true RPS, thus further demonstrating the classifier's performance. Among all the 289 RPS service, 238 are English websites, 51 are Chinese websites. To evaluate the recall of our classifier on the unlabeled dataset, We also sampled and manually verified 100 positive predictions and 100 negative predictions. As a result, 89 were confirmed to be true positives while 6 were found to be false negatives, which denotes a recall of 93.68%. We also looked into the false positive and false negative cases, in an attempt to identify directions to further enhance our classifier. Among false positives, most are websites dedicated to rating and recommending RESIP and general proxy services, e.g., https://www.bestproxyproviders.com/, and https://httpproxy.us/. Among false negatives, some were missed likely due to the fact that RESIP-relevant keywords are either buried deeply or not even available on the homepage, e.g., https://endproxies.com/ and https://rola-ip.co/. Some were missed because our dynamic crawler (a headless browser) was quickly redirected to the login pages when visiting their homepages, therefore missing the RESIP elements on their homepages. Examples include https://www.lemonproxy.net/ and https://aquaproxies.com/. We leave it as future work to conquer these performance issues.

**Key results.** In total, we have identified 399 RPS websites among which 368 are previously unknown and thus newly identified. Among these newly identified RESIP services, 64 are Chinese RESIP services and 304 are English ones.

## 3.2 Capturing Backconnect RESIPs

Given a set of Chinese RESIP services identified, we move to capture their RESIPs through multiple channels. Firstly, we adopt the infiltration framework as proposed in [25] to capture backconnect RESIPs (BC-RESIPs). *Backconnect RESIPs* are residential proxies that are hidden behind gateway servers of RESIP services. A relaying request will first be routed through the gateway server, before exiting at a RESIP. In our infiltration, 5 representative Chinese RESIP services are selected and purchased, including PinYiYun, IPIDEA, JiGuang, XiaoXiang, and Fanqie, all of which claim to have a large-scale backconnect RESIP pool. For each RESIP service, we relay network traffic, through its RESIP pool, to network servers under our control. In this case, RESIPs will serve as the ultimate exit nodes for the relayed traffic, and our servers can thus observe the public IP addresses of the involved RESIPs.

**Table 2: Results of our infiltration of China RESIP services for capturing backconnect RESIPs.**

| Provider | Period | Days | RESIPs | Probes |
|---|---|---|---|---|
| PinYiYun | 04/10/21 - 10/15/21 | 143 | 2,983,867 | 54,179,345 |
| IPIDEA | 04/10/21 - 06/09/21 | 61 | 1,375,064 | 3,599,792 |
| JiGuang | 04/10/21 - 10/15/21 | 178 | 1,732,601 | 44,057,230 |
| XiaoXiang | 04/10/21 - 10/11/21 | 177 | 1,513,640 | 57,859,909 |
| FanQie | 04/10/21 - 08/18/21 | 111 | 4,453,184 | 8,889,715 |
| Overall | 04/10/21 - 10/15/21 | 181 | 8,176,522 | 168,585,991 |

**Key results.** We carried out the infiltration spanning 6 months between April 2021 and October 2021. As a result, we have captured 8,176,522 backconnect RESIP IPs through 168,585,991 successful probes. Table 2 presents more detailed statistics in terms of the infiltration period, the captured RESIPs, and the involved probes. Besides, Figure 1 plots how the number of probes and RESIP IPs accumulate across our infiltration period.

## 3.3 Capturing Direct RESIPs

In previous works, only backconnect RESIPs were discovered and studied. However, many RESIP services identified by our classifier are observed to offer a new category of RESIPs without gateway servers as the intermediate. In another word, RESIPs in this new category will listen to specific TCP/UDP ports and accept relaying requests that are directly originated from the proxy customers. We call such a category of RESIPs as *direct RESIPs*. Leveraging this observation, we have identified, for the first time, two alternative channels for capturing RESIPs, which can help avoid the time-consuming and costly infiltration of RESIP services.

Some RESIP services have offered RESTful web APIs for customers to specify query parameters and fetch qualified direct RESIPs. These RESIP services include JiGuang, PinYiYun and XiaoXiang. We thus periodically queried these APIs during the aforementioned infiltration period between April 2021 and October 2021. Across API endpoints and RESIP services, each resulting RESIP entry usually consists of the RESIP IP and the RESIP port. Despite varied by services, a fetched RESIP may also include the credentials for access control, e.g., you need to register the IP of your host under some RESIP services before you can fetch direct RESIPs. We call direct RESIPs collected through querying APIs as *DA-RESIPs*, so as to distinguish from backconnect RESIPs as well as direct RESIPs collected through other channels. Also, we randomly sampled some freshly captured DA-RESIPs, and relayed traffic through them towards our web servers, through which, we have verified they are authentic RESIPs.

In addition, many RESIP service expose their direct RESIPs through DNS. Specifically, a RESIP service may define a set of subdomains under its control, and dynamically create and update DNS A/AAAA records to map these subdomains to active direct RESIPs at a given time. Therefore, to fetch direct RESIPs, a proxy customer can simply query the DNS resolvers for one of these subdomains. Also, RESIPs grouped under the same subdomain usually share some common properties such as ISP, location, stability, etc. For instance, *shenlongip.com* have defined 816 subdomains and 92% with each following the pattern of {*province_code*}{*city_code*}. shenlongip.com, In this case, RESIPs in different cities will be mapped to different subdomains. Leveraging passive DNS provided by our industry collaborator, we can not only identify all the up-to-date direct RESIPS offered through DNS, but extract all historical RESIP IPs. For the first time, this will allow us to carry out a temporal analysis of the RESIP ecosystem across years. Still, to distinguish from RESIPS of other categories, we define *DP-RESIPs* as RESIPs collected through passive DNS. To extensively identify services offering DP-RESIPs, we manually visited websites of all Chinese RESIP services, went through their integration documents, and verified whether direct RESIPs are available and whether they are provided through DNS. In total, 42 Chinese RESIP services are identified to offer DP-RESIPs. Also, similar to DA-RESIPs, we have verified most active DP-RESIPs are authentic ones that can relay traffic during their RESIP lifetime.

Leveraging the aforementioned two channels, we have captured 2,283,665 DA-RESIPs across three providers, as well as 1,471,361 DP-RESIPS covering 42 providers. Detailed measurements will be presented in §4 and §5 with a focus on their landscape, evolution, and potential security risks.

### 3.4 Datasets and Ethical Considerations

Before moving to measurements, we summarize available datasets among which some are generated by this study while others are either released by previous works or collected from external platforms. Furthermore, we will also summarize our ethical considerations.

*RESIP services and RESIP IPs.* Our RESIP service classifier have identified 359 RESIP services in either English or Chinese. Also, [25] has released a list of 5 RESIP services which are manually crafted and were active in 2017. Furthermore, [26] extends this dataset with 38 more services identified semi-automatically. These datasets will be used to profile the evolution of RESIP services across time, as detailed in §4. Two main datasets of RESIPs are available. One was captured by [25] in 2017 and we name it as *RESIP-2017*. The other (*RESIP-2019*) was collected in 2019 in [26]. In our study, we have identified millions of RESIPs in three categories. One is backconnect RESIPs. The other two are direct RESIPs collected from either service-specific APIs or passive DNS. A direct comparison among these datasets has been carried out and detailed results will be presented in §4.

*IP WHOIS and IP threat intelligence datasets.* IP WHOIS database contains information on assignees (owners) of IP addresses. Such information usually includes the assignee's name and type, the geolocation of the IP addresses in terms of country and city, as well as the contact information of the assignees. In our study, we utilize IPinfo [19], a popular IP intelligence service, to extract up-to-date IP WHOIS information, so as to measure the landscape and distribution of RESIPs.

To extensively profile RESIPs, we have got access to two publicly available IP threat datasets. One is VirusTotal, a well-recognized open threat intelligence platform, which aggregates malicious activities of extensive categories for various entities (e.g., files, IP addresses, domains, and URLs). Therefore, it is commonly used in security investigation and research. In our study, we queried VirusTotal with captured RESIP IPs for their maliciousness reports.

These reports are used to profile malicious activities of RESIP IPs if any, as detailed in §5. The other dataset is offered by AIWEN Tech [3], a company dedicated to profiling IP addresses. Although sharing a set of common attributes with IPinfo's IP dataset, AIWEN's dataset adds extra values since it allows us to learn more for IPs located in China, e.g., the more fine-grained geolocation information, and what threat labels an IP has been assigned with. More details about this dataset and our measurement results facilitated by this dataset will be presented in §4 and §5. In addition to the two publicly available IP threat intelligent datasets, we have also got access to a proprietary threat intelligence platform operated by our industry partner, one of the top security companies in China. Leveraging detection systems deployed nation-wide in China, this platform has aggregated fine-grained malicious traffic flows (MTFs) across a long period. In addition to the five-tuple, each MTF is also assigned with multiple valuable attributes such as the malicious categories, the timestamp, etc. Similar to VirusTotal, we queried this platform with the same set of RESIPs and utilized the resulting reports to profile malicious activities involving RESIP IPs as the clients. Still, the detailed analysis results will be presented in §5.

*Passive DNS.* In our study, passive DNS plays a critical role, especially when capturing direct RESIPs and measuring evolution of RESIP services. The passive DNS dataset adopted in our study comes from the aforementioned industry collaborator. To collect passive DNS records, our industry collaborator deploys sensors on DNS resolvers widely distributed in China. And its DNS sensors can observe an daily average of 600-billion DNS queries and responses, with the resulting passive DNS dataset covering 800-million unique domain names and 70-million unique IP addresses.

**Ethical considerations.** In our study, we have tried our best efforts to avoid any ethical issues when collecting and measuring research data. Specifically, before querying search engines for RESIP-relevant webpages, we looked into their crawling policies, and confirmed at our best judgement that our queries would not violate their policies. Also, when visiting RESIP-relevant websites, we limited our scraping scope to only in-site webpages that are referred directly or indirectly by the website landing page, and for each webpage, we crawled only once. Therefore, we believe our crawling should not have incurred any non-negligible overhead to the respective web servers. Besides, when capturing backconnect RESIPs, we sent, for each RESIP node, very few number of HTTP probes and each probe is well crafted with a small size. And we believe our infiltration shouldn't add any non-negligible traffic pressure on the RESIP devices. Also, to the best of our judgement, our industry partner collects their passive DNS dataset and threat intelligent dataset in an ethical manner, as learned from our detailed communication on this issue.

## 4 Ecosystem

Upon RESIP services and RESIP IPs of different categories, we move to characterize the RESIP ecosystem in this section. We start by measuring landscape and usage of RESIPs along with a direct comparison with previous RESIP datasets, as detailed in §4.1. Followed is §4.2 wherein we explore the potential correlations among groups of RESIPs as well as the connections among various RESIP services. Also, leveraging DP-RESIPs, we are allowed to profile the temporal evolution of the RESIP ecosystem which has distilled several interesting findings.

---

[3]https://en.ipplus360.com/home/

**Table 3: Distribution of RESIP IPs in different groups.**

| Group | IPs | /16 IPv4 | /8 IPv4 | AS | Countries | ISPs |
|---|---|---|---|---|---|---|
| BC[1] | 8.18M | 27.4K | 203 | 24.9K | 227 | 49.8K |
| F-BC[1] | 4.45M | 26.4K | 203 | 24.8K | 227 | 49.6K |
| NF-BC[2] | 3.72M | 1.2K | 48 | 17 | 19 | 272 |
| DA[2] | 2.28M | 969 | 49 | 20 | 20 | 331 |
| DP[2] | 1.47M | 2.3K | 110 | 149 | 31 | 671 |
| China | 4.66M | 3.2K | 122 | 319 | N/A | 1.1K |
| All | 9.08M | 28.4K | 208 | 24.9K | 227 | 50.3K |

[1] *BC* denotes backconnect RESIPs, *F-BC* are backconnect RESIPs captured from service Fanqie, while *NF-BC* are backconnect RESIPs from services except for Fanqie.

[2] DA denotes direct RESIPs observed from services' APIs while DP denotes direct RESIPs observed from passive DNS.

### 4.1 Landscape and Usage

As described in our methodology (§3.2), captured RESIP IPs can be grouped into three categories, depending on how they were captured. One is *backconnect RESIPs* (BC-RESIPs) captured through infiltration traffic. Then, it is direct RESIPs captured by means of querying either provider-specific web APIs or passive DNS. And they are either named as DA-RESIPs when captured through APIs or DP-RESIPs when captured from passive DNS. Next, let's take a closer look into RESIPs of these subcategories.

**Backconnect RESIPs (BC-RESIPs)**. Through a 6-month infiltration on 5 proxy providers, we have captured 8.18-million unique backconnect RESIP IPs (BC-RESIP IPs) through 169-million successful probes. Table 2 provides more details in terms of provider-specific contribution and the infiltration period. We then measured the distribution of BC-RESIP IPs in the geographic space and the network space, as listed in Table 3. These BC-RESIP IPs are widely distributed in 227 countries, 34,022 cities, 49,849 ISPs, and 203 /8 IPv4 prefixes. Such a wide distribution is a little surprising considering the involved 5 RESIP services are dedicated to serving customers in the China market. Further study reveals that Fanqie, one of the 5 services, contributes majority of non-China RESIP IPs, while the left 4 services have 99.90% of their BC-RESIP IPs located in China. This is further verified by looking into the fine-grained country distribution wherein a long-tailed pattern is observed with top 5 countries accounting for 70% of all most BC-RESIP IPs: China (46.07%), the United States (20.73%), Great Britain(4.90%), Russia (2.92%), and India (1.57%). Figure 2(a) profiles the distribution of BC-RESIP IPs across countries by means of a heatmap, which is aligned with aforementioned data points. Similar to countries, a similar long-tailed distribution is also observed for BC-RESIP IPs when measured against network blocks and ISPs, as presented in Table 10 and Table 9 (see Appendix C).

**Direct RESIPs**. In addition to backconnect RESIP IPs, we have also collected 3,240,550 direct RESIP IPs in total. Among these direct RESIPs, 2,283,665 were collected by querying APIs offered by 3 providers (XiaoXiang, PinYiYun and JiGuang), and 1,471,361 were collected through looking up passive DNS records of 42 providers

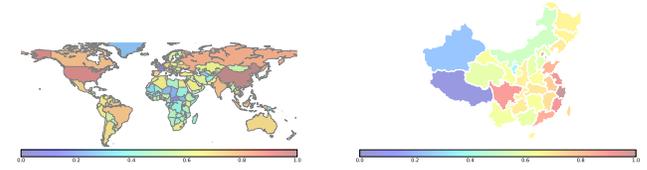

(a) The global distribution of BC-RESIP IPs.

(b) Distribution of RESIP IPs in China.

**Figure 2: Heatmaps to profile distribution of RESIP IPs in the geographic space.**

(Appendix B). Regardless of sources, these direct RESIP IPs share a similar distribution with non-Fanqie BC-RESIPs. Especially, 99.76% of direct RESIPs are located in China, and 3,240,547 are registered under 819 ISPs as revealed by IPinfo [19].

**China RESIP IPs**. Since the main focus of our study is RESIPs in China. It is thus worthwhile to move the spotlight closer to RESIPs located in China. Among all the 9.08 millions of RESIPs, 51.36% are located in China. These China RESIPs are collected from 47 proxy providers with the top 5 proxy providers being PinYiYun (67.43%), JiGuang(44.74%), XiaoXiang(32.63%), and IPIDEA(29.47%), *yunip-168.com*(4.09%) . Also, regarding network-wide distribution, China RESIP IPs have a much narrower distribution in the network space, compared to non-China ones. 99.17% China RESIP IPs are densely converged into 40 /8 IPv4 network blocks, and 88.42% are registered under 50 ISPs. Figure 2(b) profiles, by means of a heatmap, the fine-grained distribution of RESIP IPs in provinces of China. And we can see a wide coverage of all the 34 provinces. Still, a long-tailed distribution is observed with top 5 provinces have contributed 80.56% of all direct RESIPs in China: Zhejiang (35.97%), Beijing (13.08%), Jiangsu (11.98%), Fujian (10.31%), and Sichuan (9.22%).

**All RESIPs**. In total, we have collected 9,077,278 unique RESIP IPs. And these RESIP IPs are widely distributed across 227 countries and regions, 28,441 /16 IPv4 CIDRs, and 50,344 ISPs, as listed in Table 3. Among these RESIP IPs, 8,176,522 are backconnect RESIPs as observed by our multi-month infiltration, while 3,240,550 are direct ones which were extracted from provider-specific APIs as well as passive DNS datasets. The overlap of 2,339,794 RESIP IPs between backconnect and direction subsets means that some IP addresses used to serve as both.

> **Finding I**: *More than 9-million RESIP IPs have been observed in our study, constituting the largest RESIP dataset with a global distribution across 200+ countries and regions, and 50K+ ISPs.*

**A comparison with previous RESIP IP datasets**. We also compare our RESIP IP dataset with two public available ones released by previous works. One (*RESIP-2017*) was collected during 2017 by [25] through infiltrating 5 RESIP services popular at that time, while the other (*RESIP-2019*) was retrieved from 9 RESIP services in 2019 by [26]. Despite targeting regional RESIP services in China, our dataset (9,077,278 RESIP IPs in total) is even larger than either of them. Also, RESIP IPs in our dataset has a global-wide distribution that is comparable to previous datasets, in terms of countries, autonomous systems, and ISPs. Besides, only a small overlap of RESIP IPs is observed between our dataset and the previous ones, with only 299,470 RESIP IPs (3.30% of our dataset) in our dataset having ever been captured by one of the previous two datasets.

**Table 4: Statistics of the top 10 RESIP services with most direct RESIPs observed through passive DNS.**

| Provider | FQDNs[1] | RESIPs | Lifetime[2] | Total Usage | Daily Usage |
|---|---|---|---|---|---|
| yunip168.com | 636 | 191K | 1,637 | 761M | 465K |
| shenlongip.com | 816 | 181K | 948 | 1.33B | 1.41M |
| yunip520.com | 468 | 152K | 572 | 945M | 1.65M |
| vpsnb.com | 1,668 | 134K | 1,948 | 1.15B | 590K |
| jtip.in | 888 | 132K | 935 | 258M | 276K |
| xunyoull.com | 1,152 | 102K | 491 | 351M | 715K |
| 91ip.vip | 358 | 102K | 1,275 | 849M | 666K |
| upaix.cn | 2,725 | 96K | 1,863 | 324M | 174K |
| ipduoduo.cc | 309 | 90K | 2,061 | 1.34B | 648K |
| jyip.net | 2,346 | 80K | 1,792 | 819M | 457K |

[1] FQDN denotes fully qualified domains names of the respective RESIP service that are resolved to RESIPs.
[2] denotes lifetime in days the respective RESIP service in days.

More detailed overlap stats can be found in Appendix D. More importantly, our dataset has captured 4,661,934 China RESIPs, which is much larger than ones captured in previous works (148,241 in *RESIP-2017* and 183,341 in *RESIP-2019*). In summary, our dataset of RESIP IPs has added extra values compared to previous datasets in at least two aspects. One is that it can fresh our understanding of the global-wide RESIP ecosystem with 8,777,808 globally distributed RESIP IPs never captured before. More importantly, It allows us to achieve a fine-grained characterization of the RESIP ecosystem in China, which has been missed by all previous works.

> **Finding II**: *96.70% RESIPs in our dataset have never been observed in previous datasets, and 4.6-million China RESIPs are observed while it is only 148K in REISP-2017 and 183K in RESIP-2019.*

**Usage**. Given such a large scale of RESIP IPs identified, it is intuitive to wonder to what extent they have been used for traffic relaying. However, it is challenging to evaluate the usage volume of RESIPs without being an insider of RESIP services. To our best knowledge, no previous works have ever answered this question. As discussed above, many RESIP services bind RESIP IPs to their subdomains through DNS. By looking up passive DNS (pDNS), we have identified 3,635,698 relevant pDNS records and captured 1,471,361 direct RESIP IPs as the result. In addition to the mapping between a RESIP IP and a service subdomain, each of these pDNS record also contains the aggregated DNS query volume. We then utilize such DNS query volume to approximate the usage volume of RESIPs. It turns out the captured DP-RESIPs have received 16-billion DNS queries in total between 2015-06-01 and 2022-01-03. Note that, our passive DNS dataset has a limited coverage and it may not cover all DNS queries on RESIPs. Also, not every relaying request will trigger a DNS query, due to the hierarchical DNS caches. Therefore, the number of 16 billions should be considered as a lower-bound for the usage volume of RESIPs. Table 4 also presents the service-specific usage metrics of the top 10 RESIP services offering the largest number of DP-RESIP IPs. We can observe from Table 4 that RESIPs are being used frequently. Also, a service with more RESIPs observed doesn't necessarily mean it has a higher usage volume, and RESIP services of a similar scale can have very different usage volumes, as shown in the case of *yunip168.com* and *shenlongip.com*. Although *shenlongip.com* has almost the same number of RESIPs and a much shorter lifetime compared to *yunip168.com*, it has a usage volume of 1.3 billions which is almost two times as much as that of *yunip168.com*. Also,the daily usage volume of *shenlongip.com* is around 1.41 millions, the largest among all the top 10 services except *yunip520.com*

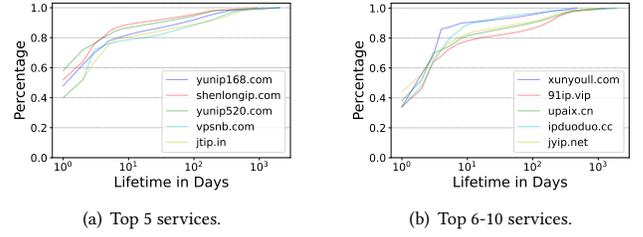

(a) Top 5 services.    (b) Top 6-10 services.

**Figure 3: The CDFs of DP-RESIPs of top 10 services over their service-specific lifetime in days.**

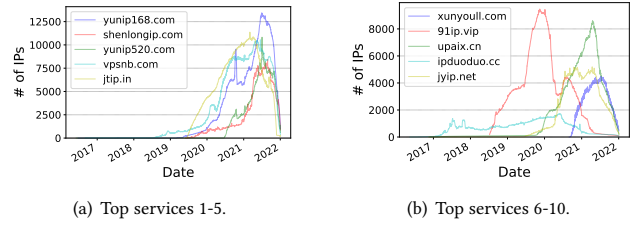

(a) Top services 1-5.    (b) Top services 6-10.

**Figure 4: Daily active DP-RESIPs across time for top services.**

## 4.2 Correlation, Lifetime, and Evolution

**Correlations among RESIP services**. We first explore the relation between direct and backconnect RESIPs. Among RESIP services under our study, PinYiYun, JiGuang and XiaoXiang provide both direct and backconnect RESIPs. It turns out that direct RESIPs and backconnect RESIPs of the same service can share a large intersection. Specifically, for RESIP service PinYiYun, its direct RESIP pool (1,277,389 RESIPs) has an overlap of 1,113,872 with the backconnect proxy pool (2,983,867 RESIP IPs), which accounts for a large portion for both pools. A similar result is also observed for RESIP service JiGuang with the overlap constituting 78.95% of the direct pool and 77.16% of the backconnect pool. In summary, a RESIP service can expose the same RESIP as both backconnect and direct.

In addition, strong correlations among proxy providers have also been observed in our study. Specifically, we profile the relation between pairs of RESIP services (e.g., a pair of $(A, B)$), by calculating the intersection rates of their RESIP IPs: $\frac{|RESIP_A \cap RESIP_B|}{|RESIP_A|}$ and $\frac{|RESIP_A \cap RESIP_B|}{|RESIP_B|}$. Among the studied RESIP services, pairs of (PinYiYun,IPIDEA) , (PinYiYun, JiGuang) and (JiGuang,PinYiYun) have shown up high intersection rate with each other. We further investigated their correlations by communicating with their respective customer services, which reveals that PinYiYun, JiGuang, and IPIDEA are controlled by the same underlying operator.

**Service-specific lifetime of RESIPs**. Among identified RESIP services, 42 RESIPs services offer direct RESIPs through dynamically resolving their subdomains to RESIP IPs. Leveraging passive DNS, we are able to learn for a DP-RESIP IP the time frame from when it shows up as an RESIP to when it disappears from the RESIP pool, in another word, the service-specific lifetime of any given

DP-RESIP. Specifically, a pDNS record will contain the active period between the first time and the last time that the DNS record is observed. Combining all these time frames will get us the lifetime of a RESIP for a specific service. Figure 3 presents such cumulative distribution of RESIPs against their service-specific lifetime in days. And we can see most DP-RESIPs (around 91%) have a short lifetime of less than 10 days, regardless of the services. Further, at least 53% RESIPs have a service-specific lifetime as short as only one day. Although Figure 3 covers only top 10 services, the observations of short lifetime is verified to be consistent across all the 42 RESIP services providing DP-RESIPs. This observation of short lifetime further highlights that RESIPs are distinct from traditional network proxies which tend to have a much longer proxy lifetime. Considering the large volume of direct RESIPs as well as the large overlap between backconnect and direct RESIPs, we believe the lifetime characteristics of DP-RESIPs can also apply to all RESIPs. Such a high churning rate of RESIPs has important security implications. Particularly, it may invalidate many well-adopted network defense techniques such as various IP blocklists, since a blocked RESIP IP will quickly be expired and attackers can easily migrate to new and clean RESIPs.

> **Finding III:** *RESIPs tend to have a short lifetime and 53% RESIPs emerge and disappear within 24 hours.*

**Evolution of RESIP services.** Similarly, passive DNS allows us to measure evolution of RESIP services in terms of the scale of their RESIP pool across time. Figure 4 presents the daily active DP-RESIPs for top 10 services between 2016-05-14 and 2022-01-03. A RESIP is considered as active on a specific date when at least one DNS records are observed on the same date to map this RESIP to a subdomain of the respective service. And we can see an obvious crest-trough pattern for most of these services. Also observed is the intensive increase of active RESIPs across all providers during 2020. Besides, it usually takes 3 years on average for a RESIP service to increase its RESIPs until arriving at the crest, in another word, gaining the maximum number of RESIPs. On the other hand, a RESIP service can collapse quickly in less than 129 days from the crest to the trough. As shown in Figure 4, all top 10 services offering DP-RESIPs have their scale of RESIPs dropped to the bottom in late 2021 and early 2022. We double checked this issue and have confirmed it is not due to any errors in data collecting and processing. An explanation we identified is that it may be subject to the crackdown of domestic crypto mining activities as conducted by the China government. China initiated this crackdown campaign in May 2021 [10], and further declared any cryptocurrency-related activities as illegal in Sep 2021 [11]. As detailed later in §5, 62.61% RESIPs have cryptojacking activities identified during 2021 and the crackdown of cryptojacking facilities and activities may have also suppressed the co-located RESIP business.

## 5 Security Risks

### 5.1 Co-Located Malicious Activities

To profile maliciousness of RESIPs, we seek help from several threat intelligence platforms, as introduced in §3.4. One is VirusTotal, a global and publicly available open threat intelligence platform, while the other is a proprietary one maintained by a security

**Table 5: Statistics of RESIP's malicious traffic flows as revealed by the proprietary threat platform.**

| RESIP Group | w MTFs[1] | ≥ 5 MTFs [2] | ≥ 10 MTFs |
|---|---|---|---|
| China | 80.05% | 68.06% | 58.79% |
| Non-China | 0.28% | 0.11% | 0.08% |

[1] denotes the ratio of RESIPs with malicious traffic flows (MTFs) observed.
[2] denotes the portion of RESIPs with each having more than 5 MTFs identified.

company dedicated to threat intelligence in China. Since the main focus of our paper is to profile the RESIP ecosystem in China, we first randomly sample 1M from RESIP IPs that are located in China and are active during 2021. Further, we randomly sample another 1M from all non-China RESIP IPs that are active during 2021, aiming to gain a global-wide understanding of RESIP's maliciousness as well as a direct comparison with previous works [25, 26]. For each RESIP IP in either group, we query both platforms for its threat reports if any.

We also observe that the aforementioned two threat platforms have different advantages in terms of profiling malicious activities relevant to a given IP. Specifically, VirusTotal focuses more on how an IP acts as a server in malicious activities, such as hosting malware or phishing websites. On the contrary, the private threat platform is found to be very helpful to profile how an IP involves in malicious activities as a client, such as being a bot due to compromise. Therefore, we present separately the maliciousness results revealed by these two platforms.

> **Finding IV:** *At least 80% China RESIP IPs have been involved in malicious traffic, while it is only 0.28% for non-China RESIPs.*

We first look into malicious traces of RESIPS as learned from the proprietary threat intelligence platform. As shown in Table 5, 80.05% China RESIP IPs have involved in at least one malicious traffic flows (MTFs) during 2021. When increasing the threshold of MTFs, a large volume of China RESIPs are still there, e.g., 68.06% have involved in at least 5 MTFs and 58.79% are of at least 10 MTFs. Contrary to China RESIPs, very few malicious traces have been observed on non-China RESIPs by the proprietary platform, with only 0.28% non-China RESIPs associated with MTFs. This is likely due to this platform's dedication to cyber threats in China.

We then take a closer look into categories of the malicious traffic flows. Table 7 presents top malicious categories along with their contribution of MTFs and the involved RESIPs. The most common malicious category is CryptoMining which have involved 62.61% RESIP IPs during 2021. Also, 11 different CryptoMining campaigns have been identified, e.g., *MiningPool* accounts for 27.27% malicious traffic flows and 35.13% RESIP IPs , and *Minerd* accounts for 16.56% malicious traffic flows and 48.30% RESIP IPs. The top list of CryptoMining campaigns along with their involvement of RESIPs can be found from Table 12. Furthermore, among RESIPs involved in MiningPool activities, around 50% were alarmed due to communications with sim.jiovt.com, a mining pool domain name that is exclusively used by a cryptomining botnet named MsraMiner [12]. Similarly, among MTFs regarded as Minerd, 72% are associated with ait.pilutce.com which turns out to be the mining server of

**Table 6: Statistics of RESIP's malicious activities as captured by VirusTotal:** *Mal* **denotes malicious.**

| RESIP Group | W Reports | Mal | W Mal URLs | W Malware |
|---|---|---|---|---|
| China | 55.98% | 1.26% | 1.25% | 0.33% |
| Non-China | 45.54% | 1.42% | 1.41% | 0.05% |

**Table 7: Top 5 categories of malicious activities involving China RESIPs.**

| Malicious Category | MTFs | % MTFs | RESIPs | % RESIPs |
|---|---|---|---|---|
| CryptoMining | 22.8M | 44.03% | 626,120 | 62.61% |
| Remote control Trojan | 13.1M | 25.32% | 569,140 | 56.91% |
| Worm | 10.4M | 20.05% | 207,577 | 20.76% |
| Botnet | 2.4M | 4.65% | 210,017 | 21.00% |
| Rogue promotion | 2.0M | 3.93% | 43,217 | 4.32% |

MTFs denotes malicious traffic flows.

a cryptojacking campaign called WannaMine [46, 47]. The second most popular malicious category is the remote control trojan , which involves 56.91% of RESIP IPs in China and accounts for 25.32% MTFs. More detailed categories of these remote control trojan activities are also listed in Appendix E.

> **Finding V**: *Among China RESIPs, 63% have ever involved in malicious cryptocurrency mining, 57% are observed in remote control trojan activities, and 21% have ever participated in botnet activities.*

We also queried Virustotal [45] with the aforementioned two groups of 1-million RESIP IPs. A VirusTotal analysis report for a given IP will include the malicious URLs the IP have ever hosted, as well as the set of malware an IP has associated with. And there are three types of associations between an IP and a malware. One is embedding wherein the IP is embedded in the payload for a given malware. The second is communicating wherein that a malware has ever communicated with the given IP. The third associate type is hosting which means the IP is found to have ever hosted the given malware for downloading. Among these three association categories, embedding and communicating are weak indicators to judge maliciousness of an RESIP IP. Instead, hosting malware is commonly recognized as a strong maliciousness indicator. Therefore, we exclude the weak indicators when analyzing VirusTotal reports, and an IP is considered as malicious only when it hosts either malicious URLs or malware.

Table 6 shows the threat statistics revealed by VirsuTotal for the two groups of randomly sampled RESIP IPs. Regardless of the RESIP group, a low rate of maliciousness is observed with only 1.26% China RESIPs and 1.42% non-China RESIPs reported as malicious, due to hosting either malware or malicious URLs. This low maliciousness ratio is consistent with previous works, e.g., [26] reports a maliciousness rate of 0.44% for RESIP IPs in cellular networks by querying VirusTotal. Despite having a similar overall maliciousness rate, a higher ratio of China RESIPs (0.33%) have associated with malware hosting, compared to non-China RESIPs (0.05%). We also take a closer look into malware or malicious URLs that are associated with RESIPS. We first analyze the categories and real-world cases of malicious URLs. Among China RESIPs, 1.25% have associated with at least one malicious URLs and these URLs are grouped by VirusTotal as three malicious categories: malware, phishing, and the general malicious. Note that an URL can be attributed with multiple categories. Among the unique 29,210 malicious URLs associated with the sampled China RESIPs, 0.40% have been labelled as phishing, 57.30% are considered as malware, and 93.89% are attributed as general malicious. Also, These URLs belong to 307 FQDNs and 120 apex domains, and top 20 apex domains account for most malicious URLs. We also manually studied top 20 apex domains and find that 15 of them are dynamic DNS services which are commonly used in web hosting at home, which also aligns with previous works. Compared to China RESIPs, a similar amount of malicious URLs are observed for non-China RESIPs.

Across China and non-China RESIPs, we have observed a large volume of URLs of the botnet Mozi. Specifically, 3,129 (0.32%) China RESIPS and 474 (0.05%) non-China IPs are found to have ever hosted Mozi payloads and facilitated Mozi infections. Mozi is a peer-to-peer botnet implemented upon the distributed hash table (DHT) [1, 4], which emerged in recent years following in the footsteps of many predecessors such as Mirai. RESIPs involving in the Mozi botnet were found to have started a http server at a randomly selected port and hosted a payload download URL in the pattern of http://ip:port/mozi.a or http://ip:port/mozi.m wherein *mozi.a* and *mozi.m* are the names of Mozi's propagation samples. For instance, *122.236.213.82*, an RESIP IP located in Shaoxing, China, has ever hosted 12 urls for Mozi payload propagation, e.g., http://122.236.213.82:50232/Mozi.a, http://mailto:info@122.236.213.82:50232/Mozi.m , and http://drolukse.rokz@122.236.213.82:50232/Mozi.m.

> **Finding VI**: *3,129 China RESIP IPs are observed to have ever hosted and distributed the infection payloads of Mozi, an emerging IoT botnet family.*

In essence, the colocation of RESIPs with various malicious activities and malware families has a strong security implications. On one hand, it can be explained that some RESIPs are harvested through malware campaigns especially botnets. On the other hand, even if RESIPs are benign, co-location with malicious programs on the same IP or even on the same device is concerning since co-located malware may be able to compromise integrity and confidentiality of any traffic relayed through the RESIPs. We leave it as our future work to further solidify the potential connections between various malware families and RESIPs.

## 5.2 Risks to the Local Networks

As revealed by previous studies, backconnect RESIPs execute by setting up one or more persistent TCP connections with the service gateways, and most backconnect RESIPs run as regular programs without privileged permissions. However, unlike backconnect RESIPs, direct RESIP programs will accept incoming relaying requests through listening to TCP/UDP ports of public IP addresses. For instance, PinYiYun's 1,275,547 direct China RESIPs all open and listen to TCP port 62456 while 53.15% XiaoXiang's direct China RESIPs have TCP port of 3000 opened for proxy traffic, as learned from online documents of these RESIP services. What is concerning is that the TCP/UDP ports opened for RESIPs can unfortunately serve as loopholes for miscreants to carry out remote attacks.

Another risk resides in the fact that sensitive organizations are found to have ever hosted RESIPs in their networks. Specifically,

AIWEN's IP dataset contains fine-gained owner information for IPs located in China and thus allows us to extract for each China RESIP IP, its respective organization information. Here, we limit our scope to several sensitive organization types including governments, enterprises, and education institutions. Specifically, 559 sensitive organizations are verified to have contributed 3,751 to the sampled 1M China RESIP IPs. Considering ethical issues and potential security risks, We decide not to reveal these organizations. We then carried out a responsible disclosure for the aforementioned 559 sensitive organizations, of which each has at least one RESIP IPs identified in their organization networks. This responsible disclosure was started by collecting organizations' contact information from various sources including IP Whois, their public websites, etc. In total, we have successfully identified at least one email address for 318 out of 559 RESIP-relevant sensitive organizations. We then sent out emails for each of them to disclose our findings related to each org as well as illustrating our research context. Also, multiple rounds of communication may be carried out to exchange more details until a concrete and deterministic response is received. By this paper submission, we have got final response from 11 out of 318 organizations, and are still communicating with, or are waiting for initial response from the left. Among these 11 organizations, 8 responded that the respective RESIPs don't belong to their networks, 1 replied that it accesses the Internet through dynamic IP addresses and thus shares IP addresses with residential users, while the left two education institutions acknowledged that it may be due to subscriptions of Internet access services by local residents that are located in their campuses. We will continue our communication with the left and will update this paper once more concrete responses are received.

> **Finding VII:** *3,232,698 China RESIP IPs have exposed at least one TCP/UDP ports to serve proxy traffic, while 3,751 China RESIPs are found to be likely located in 559 sensitive organizations including 159 education institutions, 156 goverment agencies, and 244 companies, all of which incur non-negligible security concerns.*

### 5.3 The Supply Chain of RESIPs

Given such a large volume of China RESIPs observed, it is worthwhile to explore the underlying supply chain, in another word, what channels are utilized by RESIP services to harvest RESIPs.

We first observe that many IPv4 /24 or even /16 network blocks are densely populated with China RESIPs. Specifically, 3,128 IPv4 /24 prefixes have all their child IP addresses observed as RESIPs while 10 IPv4 /16 prefixes have more than 30% of their child IPs observed as RESIPs. Such a dense distribution of China RESIPs is different from the global RESIPs revealed in previous studies. For instance, only 5 IPv4 /24 prefixes in RESIP-2017 are fully populated with RESIPs while it is only 86 in RESIP-2019. It makes us wonder whether China RESIP services have utilized any tools or services to quickly migrate their proxy nodes across IP addresses. A further investigation reveals that residential IPs may have been harvested by RESIP services through purchasing virtual private servers (VPS) from a special cloud computing service.

We name these special cloud computing services as *Switch IP in Seconds* (SIPS) since they offer a special kind of virtual private servers (VPS) that allow the customers to switch to a new residential IP in a quick and efficient manner. This is achieved by deploying each SIPS VPS in a residential building and configuring the VPS's internet access through subscribing to an ADSL service. Once an SIPS VPS is deployed, it can quickly migrate to an new IP address by repeating the logout and login operations for the subscribed ADSL service. Example SIPS services include *www.plaidc.com*, *vpsnb.com*, and *jtip.in*. One thing to note, the range of IPs that a SIPS VPS can attach to is not limited to its physical location. Instead, by means of various VPN technologies, most SIPS services allow a VPS to quickly migrate to any national-wide residential IP range as long as there is another VPS deployed in the respective IP range. Here we consider an IP address as a SIPS IP if it used to be attached to a SIPS VPS.

To further evaluate the extent to which such SIPS services have been used to provision RESIPs, we queried AIWEN's IP threat profiling dataset to decide whether a RESIP IP used to be allocated to a SIPS VPS or not. This is made possible thanks to the fact that AIWEN's IP threat dataset has a threat label to denote whether an IP address used to serve an SIPS VPS. Given the limited access granted by AIWEN, we sampled 30,000 China RESIP IPs and queried this threat dataset. As a result, 20,290 (67.63%) are labelled as SIPS IPs, which strongly indicates that SIPS services play an important role in the supply chain of RESIPs. We leave it as our future work to further profile SIPS services and their security implications.

We also explored but failed to identify the connections between RESIPs and emerging edge computing platforms, more details can be found in Appendix F.

> **Finding VIII:** *Many China RESIPs are likely harvested by means of SIPS (switch IP in seconds) services, and SIPS services offer virtual private servers that are deployed in residential networks along with the capability of quickly migrating to new residential IPs.*

## 6 Discussion

**China RESIPs versus non-China RESIPs.** Given our study's main focus is China RESIPs, we further discuss the unique characteristics of China RESIPs when compared with non-China RESIPs. Firstly, many China RESIPs have a denser distribution in the network space. Particularly, 3,128 /24 IPv4 prefixes have all of their child IPs serving as China RESIPs, and 10 /16 IPv4 prefixes have more than 30% of their IPs serving as China RESIPs. Differently, most non-China RESIPs have a more sparse distribution across network blocks. Even when combining non-China RESIPs collected in our dataset with ones from RESIP-2017 and RESIP-2019, only 111 /24 IPv4 prefixes are filled up with non-China RESIPs and 4 /16 IPv4 prefixes have more than 30% IPs serving as RESIPs.

Also, China RESIPs impose either previously unknown or higher security risks, compared to non-China RESIPs. Different from non-China RESIPs that usually serve in a backconnect mode, more than 3,232,698 China RESIPs work in a direct mode by binding to specific TCP/UDP ports and accepting external proxy traffic, which will punch a hole directly into the local networks wherein RESIP devices reside. Apparently, this can incur higher security risks compared to traditional backconnect RESIPs. Besides, the supply chain of China RESIPs looks much different from their non-China counterparts. As detailed in §5.3, RESIPs in China are likely harvested through SIPS (Switch IP in Seconds) services, which are increasingly considered as illegal along with companies offering such services hit hard by local law enforcement agencies [39]. Differently, as reported in [26] and [25], non-China RESIPs are harvested through distributing either proxy programs or proxy SDKs to residential devices, e.g., mobile phones and computers. These SIPS services provide a more efficient and convenient channel for RESIP operators to harvest residential IP resources, which raises the alarm for traditional IP-based network defense mechanisms. Also, as revealed by our industry partner's threat intelligent data, 80.05% China RESIPs have involved in various malicious activities especially cryptojacking and botnets, while it is only 0.28% for non-China RESIPs. Although such a big gap can be partially attributed to our partner's dedication to threat hunting in China, it strongly suggests the unique threat property of China RESIPs when compared with non-China ones.

**Novel contributions.** Mi, et al. [25, 26] profiled the global RESIP ecosystem with millions of RESIP IPs uncovered, hundreds of RESIP programs identified on Android and Windows platforms, as well as multiple security risks revealed. Compared to these previous works, our novel contributions can be summarized as below. First of all, we have captured more RESIP services, as well as more RESIP IPs of more categories. Particularly, we have captured 4,661,934 China RESIP IPs, which is much larger than previous works among which only 148,241 China RESIPs were captured in [25] and 183,341 were observed in [26]. More importantly, we have identified 3,240,550 direct RESIPs while all RESIPs captured in previous works are of the backconnect type. As discussed above, direct RESIPs can incur higher security risks to RESIP devices and their local network environments. Besides, leveraging our RESIP service classifier, we have identified 399 RESIP service websites, which is much larger than the 38 revealed in previous works.

Also, more security risks are uncovered in our study. Specifically, a large portion of China RESIP IPs are found to have involved in malicious network traffic especially cryptojacking and botnets. Also, it turns out that many RESIPs either share public IP addresses with or are located inside many organizations, which leads to non-negligible security risks to the involved organizations. Furthermore, many China RESIP IPs (68%) are likely harvested trough SIPS services and tools, a supply chain of RESIPs which has never been revealed before, to the best of our knowledge.

Besides, our study has also made several technical contributions. Specifically, we have built up an effective RESIP website classifier along with a set of keyword-based novel features. Also, a pipeline of collecting direct RESIPs by looking up passive DNS is designed and implemented, which can avoid the costly and time-consuming traffic infiltration when hunting RESIPs.

**RESIPs versus other network proxies.** One thing to note, RESIPs don't introduce or deploy any new network protocols. Instead, a RESIP usually shares the same set of well-defined proxy protocols with other kinds of proxies. For instance, similar to commercial VPN services, many direct RESIPs are found to have support for well-adopted VPN protocols such as PPTP and L2TP, which also explains why some RESIP services call their RESIPs as *residential VPNs*. However, RESIPs indeed have several unique characteristics when compared to other well-known network proxies, especially VPNs and the Tor network. Particularly, RESIP services usually have millions of exit nodes (RESIPs), which are far more than that of traditional proxy services, e.g., VPN services. Furthermore, traditional proxy services usually deploy their servers and exit nodes on data center networks while REISPs are located in home networks or cellular networks, which gives RESIPs incomparable advantages in terms of anonymizing or masquerading relayed traffic.

**Limitations.** There are still chances that passive DNS can be polluted with fake RESIP-relevant records, which will introduce fake direct RESIPs into our dataset. However, we argue that there are few motivations for such kinds of pollution and it would be challenging to pollute passive DNS at a large scale and across a long time. Therefore, we believe the noise in our DP-RESIP dataset can be negligible. Another limitation of our study resides in our RESIP service classifier, which may miss RESIP services for various corner scenarios as discussed in §3.1. We leave it as our future work to further improve this classifier.

**Future works.** One future work is to dive deeper into the supply chain of RESIPs, especially to understand the extent to which the RESIP ecosystem is abused by miscreants to monetize their malicious campaigns. Also, as revealed in this study, a RESIP service may not have a global visibility, but dedicate it operation to specific regions or even a single country, which leads to a set of regional RESIP ecosystems that can have unique characteristics. Therefore, another future work is to study other regional RESIP ecosystem such as RESIP services dedicated to the Russian Internet.

**Data release.** To facilitate future research, the datasets generated by this study are made publicly available on https://rpaas.site. These datasets include the RESIP IPs of different categories (e.g., backconnect and direct), the malicious traffic flows involving RESIPs, as well as the list of identified Chinese RESIP services and English RESIP services. And we believe these datasets can benefit the community in terms of reproducing our major results and building up better tools for studying RESIPs (e.g., the RESIP service classifier).

## 7 Related Works

**Web proxies.** A long line of works have looked into the security issues on web proxy services including residential proxies. Several works [42][33][24] conducted empirical studies on open web proxies to understand their usage pattern and malicious activities (e.g., traffic manipulation). Among them, Weaver et al. [48] carried out a measurement study to profile free proxy services with a focus on how they manipulate traffic. Similarly, Chung et al. [8] studied a paid proxy service to uncover content manipulation in end-to-end connections. Besides, O'Neill et al. [31] uncovered the prevalence of TLS proxies and identified thousands of malware intercepting TLS communications. Some other studies are dedicated to web proxy detection. Zhang et al. [52] proposed a proxy server detection technique by means of the distinctive characteristics of interactive traffic such as packet size and timing. Other works [21, 48] have developed techniques to detect the presence of web proxies, such as a proxy localization technique based on traceroutes of the SYN-ACK packets responding to TCP connection requests.

There are also several works on residential proxies. Mi et al. [25] reports an empirical study on five residential proxy services in the English Internet with a focus on profiling their service models and proxy IPs. Further, Mi et al. [26] studies mobile web proxies with a focus on detecting mobile proxy programs and identifying proxy traffic. Also, Akamai revealed in a white paper [44], a previously unknown attack of harvesting blackbox proxies through NAT injections. Moving forward from these works, we have uncovered for the first time, the residential proxy ecosystem in China. This ecosystem consists of 64 Chinese RESIP services, none of which have been reported or studied before. Also, to identify these RESIP services, we have designed and implemented a novel NLP-based classifier to automatically capture residential services across languages. We have also identified new techniques to capture RESIPs without relaying traffic through RESIPs. Our study has also distilled a set of novel findings such as the crest-trough evolution pattern of RESIP services, and the colocation of RESIPS with cryptomining and CDN activities.

**Cyber threat intelligence.** Several works[20, 22, 30] have built up operational systems to gather cyber threat intelligence from various sources, ranging from social networks to darknets and deepnets. Liao et al. [22] and Zhu et al. [53] mine indicators of compromise from security articles by means of NLP-augmented techniques. Besides, Khandpur et al. [20] moves spotlight to gathering threat intelligence from the social networks while Nunes et al. [30] is dedicated to hunting threat intelligence from darknets and deepnets. Bouwman et al. [7] moves forward to evaluate the coverage of two commercial threat intelligence platforms and finds out that There is almost no overlap between the two commercial platforms, nor with four large open threat intelligence feeds.

**Botnets.** Botnets have long been studied. For example, Abu et al.[2] revealed structural and behavioral features of botnets such as the high churn rate within a botnet. Collins et al. [9] studied the relationship between botnet and spamming activities. Stone et al. [40] characterized the personal data theft behavior of the Torpig botnet. In recent years, several large-scale botnet campaigns have been extensively studied, especially Mirai [3] and Hajime [17]. Our study also reveals the co-location of RESIPS with bots of several large-scale botnets including Ddostf [13], Nitol [29], and Mozi [1].

**Cryptojacking.** Due to the adoption and prosperity of cryptocurrencies, unsolicited cryptomining (cryptojacking) emerges to become an increasing cyber threat. Eskandari et al. [15] and Hong et al. [18] takes the first steps to profile in-browser cryptojacking. Naseem et al. [28] steps forward to propose a real-time detection system targeting cryptojacking. However, all these works are limited to in-browser cryptojacking, and little understanding is gained on general cryptojacking across platforms. Despite dedicated to the Chinese RESIP ecosystem, we have uncovered on residential network a large volume of cryptomining across cryptocurrencies.

**Web classification.** A long line of works have explored how to classify web pages into the predefined categories. Dumais et al. [14] has proposed the use of hierarchical structure along with support vector machine (SVM) for classifying heterogeneous web content. Besides, Sun et al.[41] explores the use of both text and context features (e.g., hyperlinks) for web classification. Furthermore, Shen et al.[38] proposes a web classification algorithm based on web summarization. Also, many works have also explored phishing webpage detection, a subtask of web classification. Liu et al. [49] and Fu et al. [16] propose the use of web page visual similarity for phishing webpage detection. Moving forward, Xiang et al.[51] presents a hybrid methodology to detect phishing websites by means of information extraction and information retrieval, while Whittaker et al. [50] presents the design and performance characteristics of a machine learning classifier deployed in Google to detect phishing web pages. Different from previous work, this classifier is learned from noisy data but still correctly classifies more than 90% of phishing pages several weeks after training concludes. In addition, Mohammad et al. [27] explores how rule-based data mining techniques can be applicable to phishing web page detection while Opara et al. [32] explores the use of deep learning into this task along with good performance achieved. Taking lessons from these website classification works especially ones on phishing webpage detection, we move forward and have pursued the task of RESIP website detection. In our detection pipeline, we utilize search engines to pre-filter irrelevant candidates, compose a representative groundtruth dataset leveraging a cycle of training and evaluation, design and select out a set of novel keyword-based features using tf-idf, and build up an effective machine learning classifier.

## 8 Concluding Remarks

In this study, we carry out the first extensive study on the China RESIP ecosystem. Our study has identified several largest-ever RESIP datasets, revealed a set of insightful findings, and raised several security concerns. Particularly, we have built up an effective RESIP website classifier as part of a pipeline to automatically capture RESIP services, which leads to the discovery of 399 RESIP services, 10 times more than ones identified in all previous works. Also, new techniques are developed to capture and monitor RESIPs in a continuous and low-cost manner. We believe the novel techniques developed, the RESIP datasets generated, and the findings distilled will benefit future research in the relevant areas.

## Acknowledgment

We would like to thank the anonymous reviewers for their insightful comments. This project is jointly supported by USTC (the start-up fund, and the Innovation Fund for Young Investigators), Shandong Provincial Natural Science Foundation (No. ZR2020MF055 , No.ZR2021LZH007, No. ZR2020LZH002, No. ZR2020QF045), and National Natural Science Foundation of China (U1836213, U19B2034).

## A  Feature Engineering for the RESIP service classifier

To identify keywords that can help distinguish RESIP and non-RESIP websites, we calculate for each keyword, its tf-idf value tf-idf$_{word, doc}$ against all the documents in the groundtruth corpus. We then average a word's tf-idf value against all RPS documents and denote it as tf-idf$_{word, RPS, AVG}$. Similarly, tf-idf$_{word, non-RPS, AVG}$ is also calculated for the word against non-RPS documents. Lastly, the gap of tf-idf tf-idf$_{word, gap}$ is defined as tf-idf$_{word, RPS, AVG}$ − tf-idf$_{word, non-RPS, AVG}$. By then, words with largest tf-idf$_{word, gap}$ values are believed to be good candidates to distinguish RPS and non-RPS documents, from which, we select out the 12 keywords as listed in Table 8.

For each keyword and the body text of each homepage, we compose the following 2 features. First, $kw_{num}$ is defined to capture the number of occurrences of a keyword in the body text. The second feature is $kw_{ratio}$ which denotes the share of the keyword accounting for the body text. $kw_{ratio}$ is thus calculated as $\frac{kw_{num}}{\# \text{ of words in the body text}}$. Then, for each keyword and one of the left 4 text components (title, keywords, description and tags), we define another binary feature, $kw_{pos}$, to capture whether the keyword exists in the text component. And $kw_{pos} = 1$ when the keyword exists in the respective text component otherwise it is 0. In total, given 12 keywords, we have crafted 72 features to distinguish RPS and non-RPS websites.

## B  RESIP Services offering direct RESIPs through DNS

Table 13 lists all the RESIP services offering direct RESIPs through DNS.

**Table 8: Keywords to construct RPS classification features.**

| Keyword | AVG tf-idf in RPS | AVG tf-idf in non-RPS |
|---|---|---|
| proxy | 0.41 | 0.01 |
| proxies | 0.40 | 0.00 |
| IP | 0.26 | 0.00 |
| residential | 0.14 | 0.00 |
| IPs | 0.11 | 0.00 |
| free | 0.08 | 0.04 |
| HTTP | 0.08 | 0.00 |
| Buy | 0.06 | 0.01 |
| price | 0.05 | 0.01 |
| rotating | 0.04 | 0.00 |
| pricing | 0.03 | 0.00 |
| provider | 0.02 | 0.00 |

**Table 9: Top /8 IPv4 prefixes contributing most backconnect RESIPs.**

| Entity | # RESIPs | % RESIPs |
|---|---|---|
| 115.0.0.0/8 | 276,007 | 3.38% |
| 183.0.0.0/8 | 250,848 | 3.07% |
| 117.0.0.0/8 | 197,085 | 2.41% |
| 113.0.0.0/8 | 193,541 | 2.37% |
| 73.0.0.0/8 | 191,296 | 2.34% |
| 49.0.0.0/8 | 190,310 | 2.33% |
| 114.0.0.0/8 | 189,258 | 2.31% |
| 180.0.0.0/8 | 169,108 | 2.07% |
| 119.0.0.0/8 | 167,625 | 2.05% |
| 42.0.0.0/8 | 163,780 | 2.00% |

**Table 10: Top ISPs with most backconnect RESIPs.**

| Entity | # RESIPs | % RESIPs |
|---|---|---|
| CHINANET jiangsu province network | 430,929 | 5.27% |
| Charter Communications Inc | 304,577 | 3.73% |
| Chinanet Jiangsu Province Network | 242,120 | 2.96% |
| China Unicom Jilin province network | 200,410 | 2.45% |
| CHINANET Guangdong province network | 192,131 | 2.35% |
| China Unicom Liaoning province network | 170,331 | 2.08% |
| AT&T Corp. | 156,653 | 1.92% |
| Comcast IP Services, L.L.C. | 153,718 | 1.88% |
| Comcast Cable Communications, Inc. | 148,019 | 1.81% |
| CHINANET-ZJ Taizhou node network | 128,559 | 1.57% |

## C  Landscape of RESIPs

Table 10 lists the top ISPs contributing the most number of backconnect RESIPs (BC-RESIPs), while Table 9 lists the top IPv4 /8 prefixes accounting for the most number of BC-RESIPs.

## D  Intersections between our RESIP dataset with previous ones.

Table 11 presents a direct comparison between our RESIP dataset and previous ones.

## E  China RESIPs' Involvement in Cryptomining and Trojans

Table 12 lists the subcategories of cryptomining and trojan activities that have involved the most RESIP IPs.

## F  Connections betwee RESIPs and Edge Computing Platforms

Through looking up Passive DNS for domain names resolved to RESIP IPs, we have found out that many China RESIP IPs have been used at the same period in CDN and edge computing, in another word, many RESIPs used to serve as edge CDN nodes or edge computing nodes. For instances, cachenode.cn, a CDN domain name, has ever been associated with 6.0% China RESIP IPs. Besides, IKuai is an edge computing platform which allows deploy computing tasks to the edge through its large-scaled edge nodes. Among all China RESIP IPs, 301,045 have also served as IKuai's edge computing nodes, accounting for 3.43% all Ikuai's nodes as identified

**Table 11: The intersection of our RESIP datasets with previous ones.**

| Group | RESIP-2017 [25] | | | RESIP-2019 [26] | | |
|---|---|---|---|---|---|---|
| | IPs | /16 IPv4 | /8 IPv4 | IPs | /16 IPv4 | /8 IPv4 |
| BC-RESIPs | 118K, 1.84%, 1.45% | 24K, 83.41%, 86.05% | 198, 100%, 98% | 222K, 2.71%, 2.72% | 25K, 84.53%, 89.30% | 199, 100%, 98% |
| DA-RESIPs | 211, 0.00%, 0.01% | 800, 2.83%, 81.97% | 49, 25%, 100% | 720, 0.01%, 0.03% | 961, 3.31%, 98.46% | 49, 25%, 100% |
| DP-RESIPs | 172, 0.00%, 0.01% | 1,830, 6.46%, 76.68% | 103, 52.02%, 93.64% | 672, 0.01%, 0.05% | 2,209, 7.62%, 94.97% | 104, 52.26%, 94.55% |
| Overall | 118K, 1.85%, 1.31% | 24k, 85.96%, 85.57% | 198, 100%, 95% | 223K, 2.72%, 2.45% | 25K, 87.63%, 89.33% | 199, 100%, 96% |

Each cell consists of three values: the overlap value, the overlap ratio with the previous dataset (the column) as the denominator, the overlap ratio with our dataset as the denominator.

**Table 12: Top subcategories of Cryptomining and Trojans involving China RESIPs.**

| Category | Subcategory | % MTFs | % RESIPs |
|---|---|---|---|
| CryptoMining | MiningPool | 27.27% | 35.13% |
| | Minerd | 16.56% | 48.30% |
| | ZombieboyMiner | 0.12% | 0.00% |
| | WannaMine | 0.05% | 0.23% |
| | SnakeMiner | 0.02% | 0.12% |
| Trojan | Generic Trojan | 18.66% | 45.03% |
| | LifeCalendarWorm | 2.57% | 13.00% |
| | Zbot | 0.89% | 2.39% |
| | DoubleGun | 0.85% | 0.19% |
| | Farfli | 0.35% | 3.22% |

by looking up passive DNS. Similarly, two edge CDN networks, namely, YunFan and CacheNode, are also identified to have their CDN nodes colocated with RESIPs on millions of IPs.

To investigate whether they are operated independent to each other, we ran on several servers IKuai's edge computing nodes for 2 months, which allows up to capture all the incoming and outgoing traffic of 1.14TB. We explore whether there are connections among RESIP services, CDN networks, and edge computing platforms. Results show that aforementioned CDN services YunFan and CacheNode are customers of IKuai. In another word, YunFan and CacheNode have deployed their CDN nodes to run on IKuai's edge computing nodes. However, we didn't observe any concrete connections between known RESIP services and IKuai. And we leave it as our future work to further investigate the supply chain of China RESIPs.

**Table 13: The list of services offering RESIPs through DNS.**

| Service | DP-RESIPs | Website Domain | Service | DP-RESIPs | Website Domain |
|---|---|---|---|---|---|
| yunip168.com | 200K | ipkuip.com | sypptp.com | 24K | 91soyun.com |
| shenlongip.com | 181K | shenlongip.com | pptpip.cn | 22K | iphai.cn |
| yunip520.com | 152K | ipkuip.com | tianqiip.com | 22K | tianqiip.com |
| vpsnb.com | 134K | ju1.kgvps.com | xjip.vip | 21K | xjip.vip |
| jtip.in | 132K | chuanyunip.com | hcnc1.com | 21K | beiked.cn |
| xunyoull.com | 102K | juip.com | 91pptp.com | 15K | runxin.run |
| 91ip.vip | 102K | 91ip.vip | pptpyun.com | 14K | ipadsl.cn |
| upaix.cn | 96K | upaix.cn | webscraping.cn | 11K | webscraping.cn |
| ipduoduo.cc | 90K | ju1.kgvps.com | cndtip.com | 6K | ipduoduo.xyz |
| jyip.net | 80K | jyip.net | guajibao.com | 5K | guajibao.cn |
| hunbo.cc | 80K | ju1.kgvps.com | beikeip.com | 4K | beikeruanjian.com |
| 236.vip | 74K | 236.vip | xfip.vip | 4K | juip.com |
| ip.vin | 71K | chuanyunip.com | qzip.vip | 2K | juip.com |
| 28289.org | 61K | ju1.kgvps.com | xgip.vip | 1K | juip.com |
| upptp.com | 52K | vps.hyidc.top | ipproxy.info | 673 | ipproxy.info |
| udp5.cn | 49K | ju1.kgvps.com | ttip.cn | 650 | jkip.com |
| leyuyun.com | 42K | leyuyun.com | gsip.vip | 556 | ipkuip.com |
| mgip.net | 35K | juip.com | xlip.vip | 506 | jkip.com |
| ipyoo.com | 29K | vps.hyidc.top | jkip.com | 316 | jkip.com |
| yoyoip.com | 27K | yayaip.com | xxip.vip | 204 | jkip.com |
| leitingip.com | 24K | ju1.kgvps.com | lyip.vip | 165 | jkip.com |

[1] A service website may become inaccessible once the respective RESIP service is shut down.